# One-wave optical phase conjugation mirror by actively coupling arbitrary light fields into a single-mode reflector

KyeoReh Lee,[1] Junsung Lee,[2] Jung-Hoon Park,[1,*] Ji-Ho Park,[2] and YongKeun Park[1,†]

[1]*Department of Physics, Korea Advanced Institute of Science and Technology, Daejeon 305-701, Republic of Korea.*
[2]*Department of Bio and Brain Engineering, Korea Advanced Institute of Science and Technology, Daejeon 305-701, Republic of Korea.*

Rewinding the arrow of time via phase conjugation is an intriguing phenomena made possible by the wave property of light. To exploit this phenomenon, diverse research fields have pursed the realization of an ideal phase conjugation mirror, but an optical system that requires a single-input and a single-output beam, like 'natural' conventional mirrors has never been demonstrated. Here, we demonstrate the realization of a one-wave optical phase conjugation mirror using a spatial light modulator. An adaptable single-mode filter is created, and a phase-conjugate beam is then prepared by reverse propagation through this filter. Our method is simple, alignment free, and fast while allowing high power throughput in the time reversed wave, which have not been simultaneously demonstrated before. Using our method, we demonstrate high throughput full-field light delivery through highly scattering biological tissue and multimode fibers, even for quantum dot fluorescence.

The light reflected from mirrors obeys the law of reflection. This limits the usage of such mirrors to simple optical components can only redirect propagating light in a specific direction. However, if the phase relationship on the mirror surface can be properly controlled, one could imagine unconventional mirrors with various different consequences.

One of the most popular types of unconventional mirror is the phase conjugation mirror (PCM) [1]. The PCM conjugates the phase on the mirror surface, and causes the reflected light to rewind through the path that input light have travelled. This 'time-reversal' property of PCMs has been tried in various optical applications such as aberration canceling [2,3], pulse compression [4,5], laser resonators [6,7], holography [8], and suppression of multiple light scattering in biological tissues [9-14]. More direct demonstrations have also been performed in using acoustic waves [15] and microwaves [16], where conventional electronics can directly measure, and generate the phase profile in real time.

Historically, optical PCMs have been usually aided by several non-linear optical effects such as, stimulated Brillouin backscattering [17], four-wave mixing [18], and the photorefractive effect [19]. The functionality of the non-linear PCMs rely on the interferences between input lights. Absence of electronics allows non-linear PCMs to handle large degree of optical information even in real-time. However, simultaneously, the non-linearity raises technical difficulties that limit the versatility of the PCMs; e.g., low reflectivity, signal sensitivity, as well as unwanted non-linear effects.

Inspired from the development of electronics, digital optical phase conjugation (DOPC) has recently been suggested to dodge the non-linear disadvantages [12,13,20]. The DOPC measures the input field by interferometry, and then the conjugated field is generated from the other arm with a spatial light modulator (SLM). The idea of separation of input/output arms is especially beneficial to the amplification of conjugated light; the output light can be freely amplified regardless of the input light intensity. However, for perform the conjugation, the desired optical field should be exactly regenerated by the SLM, which requires extremely precise alignment and calibration between the camera and the SLM in six degrees of freedom [21]. Therefore, ambient perturbations and noise can easily disable the functionality of DOPC. This vulnerability have severely discourage the practical use of DOPC.

The ideal PCM will be an optical system that requires a single input beam, like conventional mirrors (Fig. 1a), as is frequently described in conceptual figures (Fig. 1b). Unfortunately, all previously demonstrated PCMs rely on separate beam paths for the input and output beams (Fig. 1c). Therefore, the requirement for the utilization of reference light and the exact 'correlation' between the input light field and the phase conjugated output beams generates all the critical issues described above.

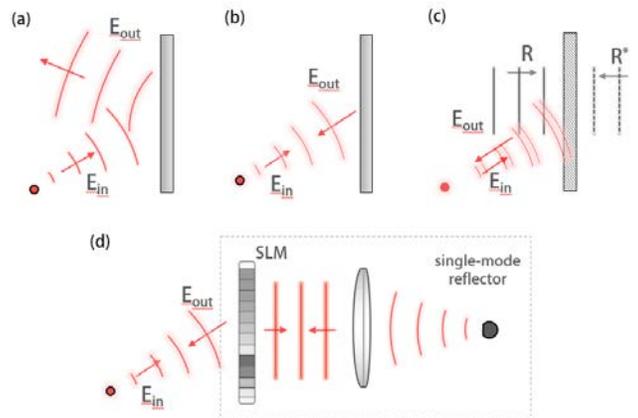

FIG. 1 (color online). Mirrors as optical systems. (a) Natural mirror system. (b) Ideal PCM system requires single light input, $E_{in}$. (c) Conventional PCMs requires additional reference lights, $R$ and $R^*$. (d) Proposed idea utilizing a wavefront shaper and a single-mode reflector.

Here, we present optical phase conjugation mirror with a one-wave configuration for both input and phase conjugated beams [Fig. 1(d)]. We propose and experimentally demonstrate naturally occurring phase conjugation by actively coupling arbitrary light fields into a single-mode reflector. A single-mode channel can be any single phase translating optical system such as a diffraction limited diaphragm or a single-mode waveguide. By definition, single-mode channels can only carry a single phase, so the light that propagates in the opposite direction through the channel will naturally be its phase conjugated counterpart. Exploiting this relationship, we couple arbitrary input light into the single-mode channel using a wavefront shaper such as a SLM, a dynamic mirror device, or a deformable mirror. Thereafter, the reflected light from the single-mode channel will follow the exact time-reversed trajectory of the input light. We can also couple another light source with the same frequency onto the channel to amplify the output light intensity, rather than using the direct reflection of the input light.

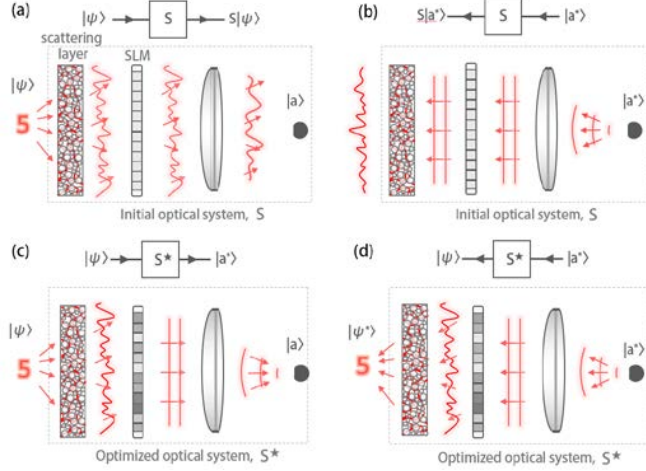

FIG. 2 (color online). Principle of the one wave phase conjugation mirror. (a) For the initial optical system $\mathbf{S}$, the input light $|\psi\rangle$ is scrambled and the light is randomly distributed over multiple channels. (b) The reflected light from the single-mode channel $|a^*\rangle$ will generate another randomly distributed output. (c) For the optimized optical system $\mathbf{S}^\star$, the SLM actively couples arbitrary input fields into the single-mode channel. (d) In this case, the reflected single mode emitter would time-reverse the incoming pathway.

Let an arbitrary optical system be described as a scattering matrix $\mathbf{S}$. Then, for the given input field ($|\psi\rangle$), the output field can be formulated as,

$$\mathbf{S}|\psi\rangle = \alpha|a\rangle + \sum_n \beta_n |a_n^\perp\rangle, \qquad (1)$$

where $|a\rangle$ is a selected single-mode channel, and $|a_n^\perp\rangle$ are a basis of perpendicular space with respect to $|a\rangle$ [Fig. 2(a)]. Ignoring the light loss or gain in the system, $\alpha$ and $\beta_n$ are arbitrary complex coefficients that satisfy $|\alpha|^2 + \sum_n |\beta_n|^2 = 1$. For any arbitrary light (input channel), the proposed system can actually conjugate the phase of light that reaches the desired channel [Fig. 2(b)] even prior to wavefront shaping,

$$\mathbf{S}|a^*\rangle = \gamma|\psi^*\rangle + \sum_n \delta_n |\psi_n^{*\perp}\rangle. \qquad (2)$$

However, in this case, $|\gamma|^2 \approx |\delta_1|^2 \approx |\delta_2|^2 \approx \cdots \approx |\delta_n|^2 \approx \cdots$, resulting in a speckled output beam. Here, according to the nature of reciprocity [22], the portion of output energy allocated to the conjugated initial field ($|\gamma|^2$) is identical to the portion of input energy allocated to the desired single channel ($|\alpha|^2$),

$$|\gamma|^2 = |\langle\psi^*|\mathbf{S}|a^*\rangle|^2 = |\langle a|\mathbf{S}|\psi\rangle|^2 = |\alpha|^2. \qquad (3)$$

Therefore, by introducing an appropriate phase mask onto the SLM which is incorporated into the scattering matrix $\mathbf{S}$, we can obtain the effective scattering matrix $\mathbf{S}^\star$ which focuses the energy onto the channel $|a\rangle$ and also results in the efficient phase conjugation of the input beam [Figs. 2(c) and 2(d)]. In this reconfigured case, light reflection is anti-causal system.

We applied the parallel optimization algorithm proposed by Cui [23], but the proposed method is not limited to this specific algorithm; any wavefront shaping algorithm can also be employed [24,25]. We divided the SLM pixels into two groups in a checkerboard manner to supply reference fields for each other. For each group, we measured $4T$ intensity points, $x = 0, 1, 2, \ldots, 4T-1$, where $T$ is the pixel number of the group.

During the measurement, we modulated each pixel according to unique frequencies from $(T+1)/4T$ to $1/2$ with $1/4T$ frequency increment. Then, the field generated on the single channel ($E_p$) is as follows:

$$E_p(x) = E_0 + \sum_{k=1}^{T} t_k e^{i2\pi \frac{T+k}{4T} x}, \quad (4)$$

where $E_0$ is the reference field generated by the other group, and $t_k$ is the field generated by $k$-th pixel of the group. Therefore, the measured intensity is as follows:

$$I_p(x) = E_0^* E_0 + E_0^* \left( \sum_{k=1}^{T} t_k e^{i2\pi \frac{T+k}{4T} x} \right)$$
$$+ E_0 \left( \sum_{k=1}^{T} t_k^* e^{-i2\pi \frac{T+k}{4T} x} \right) + \left( \sum_{k,l=1}^{T} t_l^* t_k e^{i2\pi \frac{k-l}{4T} x} \right). \quad (5)$$

Since the last interference term (due to beating) only exists (-1/4, 1/4) in the frequency domain, we can simply measure the $E_0^* t_k$ by taking the Fourier transform of the measured intensity and the taking the information in $[(T+1)/4T, 1/2]$ in the frequency domain. To maximize the intensity, we set

$$\phi_k^{opt} = -\angle(E_0^* t_k) \quad (6)$$

for each SLM pixel. After the individual optimizations of the two groups, we matched the global phase between the groups by shifting the global phase of one group to match the other group [26].

To demonstrate the capability of the proposed method, we obtained full-field optical phase conjugation through highly scattering biological tissues. Biological tissues produce multiple scattering, which makes it extremely challenging to deliver optical information through the tissue. However, in spite of the complexity, it is still possible to reconstruct the input light through phase conjugation because light transport in biological tissue is almost lossless and linear in the optical regime. The concept of field reconstruction is identical to that of aberration correction, but the difference is that the much higher degree of scattering makes the light trajectories inside the tissue greatly sensitive to the input light. Thus, to achieve 'time-reversal' for biological tissue, extremely meticulous phase conjugation is required.

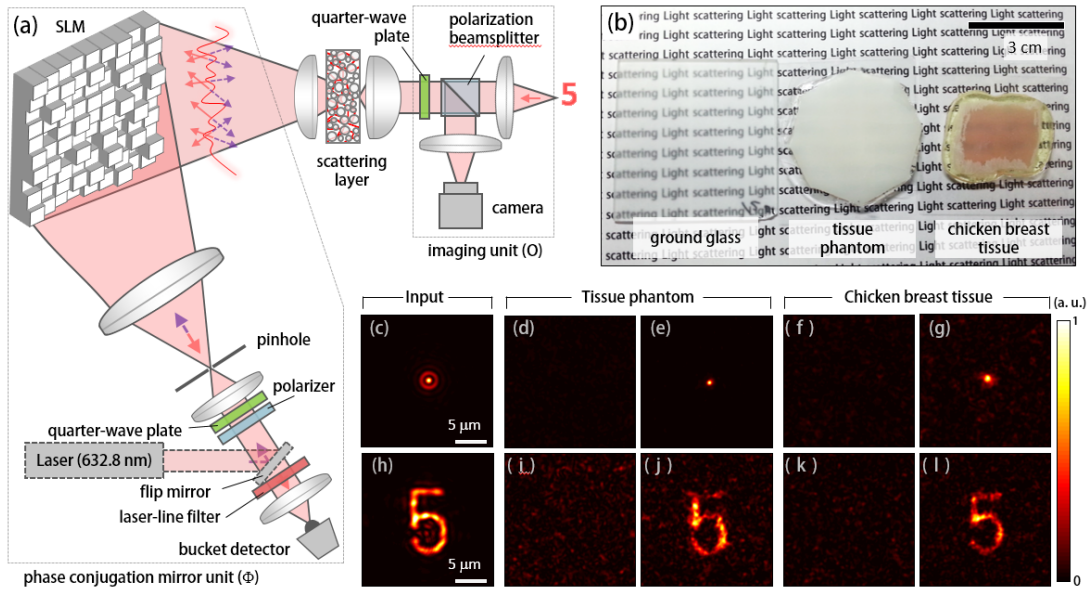

FIG. 3 (color online). Suppression of multiple scattering using the phase conjugation mirror. (a) Schematic of the optical setup. Red arrows represent the input wavefront, dotted purple arrows represent the time-reversed wavefront. SLM, spatial light modulator. (b) Scattering mediums: ground glass (120 grit, displayed for the comparison), tissue phantom, and chicken breast tissue. (c), (h) Input images. Initial scrambled field images by (d), (i) tissue phantom and (f), (k) chicken breast tissue. Reconstructed images created behind of (e), (j) tissue phantom, and (g), (l) chicken breast tissue.

Figure 3(a) shows the schematic of the experimental setup. Input images were generated by masking laser light with a 30-μm-diameter pinhole or a negative USAF target (the letter '5' in group 1). We used a tissue, and chicken breast tissue [Fig. 3(b)] as highly scattering media [27]. Our single-mode channel was a pinhole with a diameter of 10 μm, which corresponded to the sub-diffraction limited spot size of the relay optics. A quarter-wave plate and a polarizer were used to achieve any general polarization state for the single-mode channel. Then, the detection polarization was adjusted to maximize the initial intensity read by the bucket detector to achieve the best quality phase conjugation. To block ambient light, we placed a laser-line filter in front of the bucket detector. The average total optimization time took 9.5 seconds, and the average enhancement factor was 620, using a deformable mirror with 1014 modes.

After the optimization, the flip mirror was flipped up to direct the laser in the opposite direction through the pinhole, and the single-mode channel now emitted light. The phase conjugated light could therefore be amplified arbitrarily using an independent light source. To observe the reconstructed conjugated images, we rotated the quarter-wave plate in the imaging unit (O), because the conjugated output light is identically polarized to the polarization of the input light. After the phase conjugated images were taken, the corresponding initial output images were also taken in the identical situation but with a flat wavefront displayed on the SLM. For both scattering mediums, the phase conjugated images were successfully reconstructed [Figs. 3(c)-3(l)].

Another promising application for PCMs is its adaptation to multi-mode fibers (MMFs), which particularly important for applications in biomedical endoscopic and spatially multiplexed optical communication. However, due to the modal dispersion of MMFs, the information distributed at the input of the MMF becomes scrambled at the output making direct information transfer difficult. PCMs have been proposed as an effective method to cancel the dispersion of MMFs [2,28,29]. Here, we demonstrate that the present alignment-free optical phase conjugation mirror can effectively undo the modal dispersion of a commercial MMF, and successfully recover the input image through phase conjugation [Figs. 4(a)-4(d)].

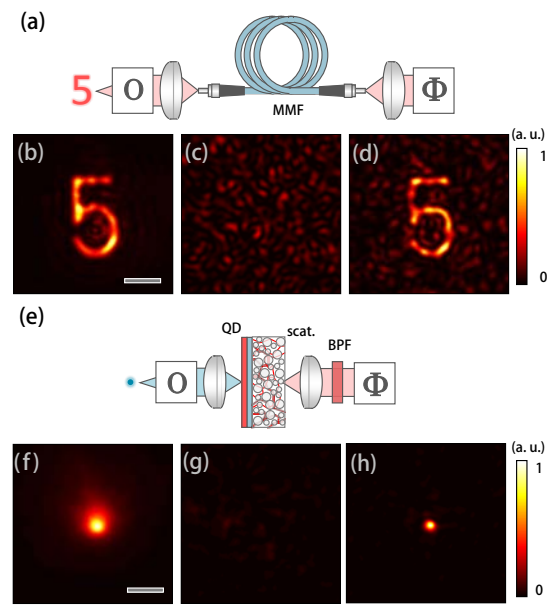

FIG. 4 (color online). Demonstration of various PCM applications. (a), (e) Schematic of the optical setups for delivering images through MMF, and for QD fluorescence phase conjugation through scattering media, respectively. Please note that the symbolized units ('O' and 'Φ') are identical to Fig. 3(a). (b), (f) Input field images. Scale bar: 30 μm and 5 μm, respectively. (c), (g) Initial scrambled images. (d), (h) Reconstructed images.

An important factor regarding the utilization of our technique is its applicability to incoherent light sources [30]. For example, quantum dots (QDs) or fluorescent proteins can be functionally tagged to specific targets inside living tissue and act as internal beacons using phase conjugation in a totally non-invasive manner. Achieving robust phase conjugation for such sources will be a crucial step in expanding our technique to biomedical applications.

To verify such characteristics, we excited dried QD with a 488 nm laser diode [Fig. 4(e), SI]. The fluorescence emitted from spatially distinguishable points did not interfere with each other due to the temporal incoherency of fluorescence, and incoherence between the QDs. The incoherence severely reduces interference contrast. To minimize the issue, we used point-like input images [27]. In addition, given that the intensity was much lower than in the previous applications, thinner chicken breast tissue (410 μm) was used as the scattering medium, and more time (36 s) was taken for the optimization.

The results are shown in Fig. 4(g)-(h). Using the present method, quantum dot fluorescence, exited by a point illumination of the 488 nm laser diode [Fig. 4(f)], can be focused at the sample position without image distortion after double passing the highly scattering chicken breast tissue. In contrast, simple reflection only generates highly dispersed weak speckle patterns [Fig. 4(h)].

In sum, we proposed and experimentally demonstrated one-wave optical phase conjugation mirror, a time-reversal reflection module that is self-aligning. We expect immediate and diverse applications of the present method including focusing and imaging inside biological tissues [25,31-33], near-field control [34,35], nonlinear optics [36], optical fiber communications [37], stable laser resonators [38], and delivering high-power laser beams through turbulent atmosphere [39]. In principle, the technique is also applicable for any form of waves including infrared, ultraviolet, seismic waves, and even X-rays. We hope that this easily distributable technique will immediately find diverse usage in the many research areas where wave transport through aberrating or scattering volumes is inevitable.

This work was supported by the Korean Ministry of Education, Science and Technology, and the National Research Foundation (2014K1A3A1A09063027, 2013M3C1A3063046, 2012-M3C1A1-048860, 2014M3C1A3052537).

*Current affiliation: Howard Hughes Medical Institute, Janelia Research Campus, Ashburn, VA USA
†yk.park@kaist.ac.kr